\documentstyle[aps,epsf]{revtex}

\draft
\title{Dynamics and thermodynamics of the \\spherical frustrated Blume-Emery-Griffiths model}
\author{A. Caiazzo, A. Coniglio, M. Nicodemi}
\address{Dipartimento di Scienze Fisiche, INFM, Unita' di Napoli,\\ Monte Sant'Angelo, I-80126 Napoli, Italy}
\begin{document}
\maketitle

\begin{abstract}
We introduce a spherical version of the frustrated Blume-Emery-Griffiths model and solve exactly the statics and the Langevin dynamics for zero particle-particle interaction $\left( K=0\right) $. In this case the model exhibits an equilibrium transition from a disordered to a spin glass phase which is always continuous for nonzero temperature. The same phase diagram results from the study of the dynamics. Furthermore, we notice the existence of a nonequilibrium time regime in a region of the disordered phase, characterized by aging as occurs in the glassy phase. Due to a finite equilibration time, the system displays in this region the pattern of interrupted aging.
\end{abstract}
\section{Introduction}

Many important features of spin glass models at mean field level have come
out by studying their relaxional Langevin dynamics from a random initial
condition\cite{SZ,KT,CDP,CHS,CK,CK1,CD,CLD,BCKM}. The structure of the
dynamical equations for the correlation and response functions has reaveled
some analogies with other types of complex systems in which the disorder is
apriori absent: at equilibrium the dynamics becomes formally identical to
the Mode Coupling Theory (MCT), which is the main approach to the
supercooled liquids near the glass structural transition \cite{G}. Thus
there have been strong feelings that the two types of systems are deeply
connected; in the glasses an effective disorder is self-induced by the slow
dynamics of the microscopical variables \cite{KT}.

For spin glass systems, the dynamical equations have been studied also in
the low temperature phase \cite{CK,CK1,CD,CLD,BCKM}. These works provide a
suggestion to extend the MCT below the glass temperature \cite{BCKM1}. One
of the main result has been that for these temperatures the system never
reaches the equilibrium, but rather displays an off-equilibrium behaviour
where the dynamics depends on the whole history of the system up to the
beginning of its observation and is characterized by the loss of validity of
typical equilibrium properties, as the time traslational invariance (TTI)
and the fluctuation-dissipation theorem (FDT). One can thus establish
contact with some nonequilibrium experimental observations, namely the slow
relaxations and the aging phenomena which are observed for real spin glasses
and many other complex systems \cite{BCKM}.

Despite the cited resemblance, spin glasses are microscopically quite
different from liquids and thus not suitable to their description. Recently,
to make stronger connections with liquids, some models have been introduced
which combine features of spin glasses and lattice gas. Being constituted of
particles, they allow to introduce the density and other related quantities
which are usually important in the study of liquids. In this regard we
consider the frustrated Blume-Emery-Griffiths (BEG) model \cite{SNA,CL}, which
is a quite general framework to describe different glassy systems. Its mean
field hamiltonian is: 
\begin{equation}
H=-\sum_{i<j}J_{ij}s_in_is_jn_j-\frac KN\sum_{i<j}n_in_j-\mu \sum_in_i
\label{BEG ham}
\end{equation}
where $s_i=\pm 1$, $n_i=0,1$, $\mu $ is the chemical potential and $J_{ij}$ are quenched gaussian interactions
having zero mean and variance $\left[ J_{ij}^2\right] _J=1/N$ \cite{NO}. Essentially the model consists of a lattice gas in a
frustrated medium where the particles have an internal degree of freedom,
given by their spin, which may account, as an example, of the possible
orientations of complex molecules in glass forming systems. These steric
effects are indeed greatly responsible for the geometric frustration
appearing in glass forming systems at low temperatures or high densities.
Besides that, the particles interact through a potential depending on the
coupling $K$. In particular for $K=0$ one recovers the Ghatak-Sherrington
model \cite{GS} and for $K=-1$ the Ising Frustrated Lattice Gas model \cite
{ANS}; this last case is related to the problem of the site frustrated percolation \cite{C} and has been also used in the presence of gravity to describe granular materials \cite{GM}. However, as found by standard replica theory \cite{SNA,CL}, the
model does not display substantial differences by varying $K$. The phase diagram in the plane $\mu -T$ shows a
critical line separating a spin glass phase from a disordered one; the
transition is continuous for large $\mu $, as for the standard Ising spin
glass ($n_i=1$), up to a given value $\mu ^{*}_{K} $ below which becomes
discontinuous; in this region the Parisi solution has been obtained only recently in \cite{CL}. Moreover, we note that a dynamical treatment of the model is still
lacking.

We propose a spherical version of the frustrated BEG model which allows a complete
analysis of its equilibrium properties and even of Langevin dynamics. In
this paper we study this model for $K=0$ leaving the general case to future
investigations \cite{FW}. We find an equilibrium transition to a spin glass phase for $\mu \geq -1$, which
is always continuous, like in the spherical SK model \cite{KTJ}, except for $%
\mu =-1$ and $T=0$ (see Fig. 1). Furthermore, we investigate the Langevin
dynamics of the various two-time functions and of density in the whole phase
diagram. We get exact expressions of these quantities that are in general
rather complex; they  depend on several characteristic time scales whose number
changes in the various regions of the phase diagram. In particular, the
largest of these times is found to represent the charactristic equilibration
time of the system, $\tau _{eq}$. In the nonglassy phase this is finite and
for waiting times $t^{\prime }>\tau _{eq}$ the two-time functions obey TTI
and FDT. This is no more true for $t^{\prime }<\tau _{eq}$, the system being
still out of equilibrium. By studying this regime near the critical line
where $\tau _{eq}$ is large, we find two different behaviours of the systems
for $\mu >-1$ and $\mu \simeq -1$. We give here a first qualitative
description of them. In the case $\mu >-1$ FDT still holds for $t-t^{\prime
}\ll t^{\prime }$, so that deviations from the equilibrium case occur only
for very small values of the correlation. Instead, as $\mu $ is close to $-1$%
, the range of correlation values in the FDT regime decreases and for $\mu
=-1$ this vanishes; if $t,t^{\prime }$ are sufficiently lower than $\tau
_{eq}$ the correlation function scales as a power of $t^{\prime }/
t$. Thus, in this region the observation of the
system over short time scales could wrongly lead to conclude that the
system is in the glassy phase. Then, for a large but finite $\tau _{eq}$, the model follows the pattern
of interrupted aging. Finally, $\tau _{eq}$ diverges in the whole
glassy phase and the system displays essentially the same nonequilibrium
behaviour of the spherical SK model \cite{CD}.

The paper is organized as follows. In Sec. II we define the spherical frustrated BEG
model. In Sec. III we deal with the statics using the theory of large
random matrices and analyse the thermodynamical properties. In Sec.
IV we introduce the Langevin dynamical model and the various quantities of
interest. In Sec. V we solve the integral equation related to the spherical
constraints and compute the density. In Sec. VI we discuss the dynamics in
the disordered phase and, after the identification of the equilibration time 
$\tau _{eq}$, we analyse the equilibrium regime $\left( t^{\prime }>\tau
_{eq}\right) $ and the out of equilibrium one $\left( t^{\prime }<\tau
_{eq}\right) $. In Sec. VII we consider the nonequilibrium dynamics in the
glassy phase $\left( \tau _{eq}=\infty \right) $. Finally, in Sec. VIII we
present some conclusions. Furthermore, in the App. A and B we study in
detail respectively the equilibrium saddle point equation and the dynamics
when this equation has degenere roots.

\section{Definition of the model}

First of all we note that the hamiltonian (\ref{BEG ham}) can be rewritten
in terms of two new Ising spin fields $s_{1i},s_{2i}=\pm 1$: 
\begin{equation}
H=-\sum_{i<j}\frac{J_{ij}}4\left( s_{1i}+s_{2i}\right) \left(
s_{1j}+s_{2j}\right) -\frac K{4N}\sum_{i<j}\left( s_{1i}s_{2i}+1\right)
\left( s_{1j}s_{2j}+1\right) -\frac \mu 2\sum_i\left( s_{1i}s_{2i}+1\right)
\label{BEGs ham}
\end{equation}
where: 
\begin{equation}
\left\{ 
\begin{array}{l}
s_{1i}=s_i \\ 
s_{2i}=s_i\left( 2n_i-1\right)
\end{array}
\right. \quad \Leftrightarrow \quad \left\{ 
\begin{array}{l}
s_i=s_{1i} \\ 
n_i=\frac 12\left( s_{1i}s_{2i}+1\right)
\end{array}
\right.
\label{cv}
\end{equation}
\cite{NO}. The
4-fields interaction in (\ref{BEG ham}) is replaced in (\ref{BEGs ham}) by
four double field interactions; furthermore the hamiltonian (\ref{BEGs ham})
is symmetric under the exchange of the two spin fields. The overlap $%
q=\left[ \left\langle s_in_i\right\rangle ^2\right] _J$ and the density $d=\left[
\left\langle n_i\right\rangle \right] _J$, which are two usual order
parameters for a diluted spin glass, now become respectively $\frac 14\left[
\left\langle s_{1i}+s_{2i}\right\rangle ^2\right] _J$ and $\frac 12\left[
\left\langle s_{1i}s_{2i}\right\rangle +1\right] _J$. So far we have just
reformulated the BEG model. Now to define our spherical version let the
Ising constraints fall in (\ref{BEGs ham}) and replace them by the spherical
ones: $\sum_is_{1i}^2=\sum_is_{2i}^2=N$. This particular choice of the
variables to sphericize aims to obtain an exactly solvable model. It
recovers the spherical SK model \cite{KTJ} in the large $\mu $-limit ($%
s_{1i}=s_{2i}$). Below we'll consider the case $K=0$.

To study the model it is convenient to diagonalize the interaction matrix $%
J_{ij}$ ($\sum_jJ_{ij}\eta _{j\lambda }=\lambda \eta _{i\lambda }$) and work
with the variables $s_{a\lambda }=\sum_i\eta _{i\lambda }s_{ai}\quad (a=1,2)$%
; these obey the properties $\sum_is_{ai}s_{bi}=\sum_\lambda s_{a\lambda
}s_{b\lambda }$ and $\sum_{ij}J_{ij}s_{ai}s_{bj}=\sum_\lambda \lambda
s_{a\lambda }s_{b\lambda }$. In the $N\rightarrow \infty $ limit the
eigenvalue density $\rho \left( \lambda \right) $ satisfies the Wigner
semi-circle law: 
\begin{equation}
\rho \left( \lambda \right) =\left\{ 
\begin{array}{ll}
\frac 1{2\pi }\sqrt{4-\lambda ^2} & \left| \lambda \right| <2 \\ 
0 & \left| \lambda \right| \geq 2
\end{array}
\right.
\end{equation}
the quantities $\eta _{i\lambda }$ are gaussian variables with zero mean and
moments $\left[ \eta _{i\lambda }^{2k}\right] _J=\frac{\left( 2k-1\right) !!%
}{N^k}$; they are uncorrelated to the eigenvalues and among themselves apart
the orthonormality and closure conditions.

\section{Equilibrium properties}

\subsection{Saddle point equation}

The statics can be solved by standard techniques for spherical models and
the above properties of large random matrices \cite{KTJ}. In the $%
N\rightarrow \infty $ limit the free energy is evaluated by steepest
descent, by imposing saddle point conditions with respect to the two
Lagrange multipliers $z_1$ and $z_2$, introduced by the spherical
constraints. These equations just reproduce the two constraints satisfied on
average, $\sum_\lambda \left\langle s_{a\lambda }^2\right\rangle =N\quad
(a=1,2)$. Explicitly they are reduced to the only one: 
\begin{equation}
\frac 1{2\beta ^2}\left[ z-\sqrt{z^2-4\beta ^2}\right] =1-\frac 1{z+2\beta
\mu }  \label{saddle-eq}
\end{equation}
where $z=z_1+\beta \mu =z_2+\beta \mu $ has to be greater than the two branch
points $\left\{ 2\beta ,-2\beta \mu \right\} $. This condition is satisfied
by a unique solution of (\ref{saddle-eq}), denoted by $z_s$, for each $T$ if 
$\mu <-1$ and above a critical line, $T=T_c\left( \mu \right) $, for $\mu
\geq -1$. This region of the phase diagram identifies the disordered phase
(labelled by P in Fig. 1). The critical line is located by $z_s$ reaching
the branch point $2\beta $ and is given by: 
\begin{equation}
T_c\left( \mu \right) =\frac{\mu +1}{\mu +3/2}  \label{crit.line}
\end{equation}

A detailed study of Eq. (\ref{saddle-eq}) is given in App. A. Below the
critical line (phase SG in Fig. 1) this equation is not satisfied for $%
z>2\beta $. The equilibrium saddle value of $z$ sticks at the branch point $%
2\beta $, and to preserve the spherical constraints a spontaneous
magnetization arises along the eigenvector with eigenvalue $2$. Actually,
the diluted overlap $q=\frac 14\left[ \left\langle
s_{1i}+s_{2i}\right\rangle ^2\right] _J$ is found to vanish when Eq. (\ref
{saddle-eq}) holds and becomes just $\frac{\left\langle s_{a,\lambda =2}\right\rangle
^2}N$ below the critical line, i.e.: 
\begin{equation}
q=1-\frac 1\beta -\frac 1{2\left( \beta +\beta \mu \right) }=1-\frac T{%
T_c\left( \mu \right) }  \label{q}
\end{equation}
The transition at the line $T=T_c\left( \mu \right) $ is continuous for each 
$\mu >-1$ and discontinuous in the point $(\mu =-1,T=0)$. Indeed the zero
temperature value of $q$ is $\vartheta \left( \mu +1\right) $ with $q=0$ for 
$\mu =-1$.

Note that model could be solved using the replica trick, where a replica
symmetric ansatz yields identical results.
\begin{figure}[h]
\begin{center}
\epsfxsize=7cm
\epsffile{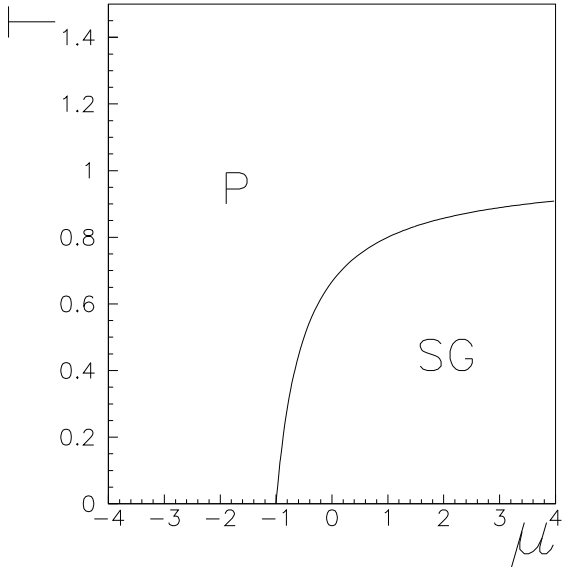}
\caption{The phase diagram $T$ vs. $\mu $. The critical line separating the
two phases is given by Eq. (\ref{crit.line}) for $\mu \geq -1$. The transition
is always continuous except for the point $\left( \mu =-1,T=0\right) $.}
\end{center}
\end{figure}

\subsection{Free energy and related quantities}

The free energy per site $f$ can be explicitly evaluated: 
\begin{equation}
\beta f=\left\{ 
\begin{array}{ll}
-\frac{z_s+2\beta \mu }2-\ln 4\pi +\frac 12\ln \left( z_s+2\beta \mu \right)
+\frac{\beta ^2}4\left( 1-\frac 1{z_s+2\beta \mu }\right) ^2-\frac 12\ln
\left( 1-\frac 1{z_s+2\beta \mu }\right) & P \\ 
-\left( \beta +\beta \mu \right) -\ln 4\pi +\frac 12\ln 2\left( \beta +\beta
\mu \right) +\frac 14+\frac 12\ln \beta & SG
\end{array}
\right.
\end{equation}
it corresponds to a negative low temperature entropy which diverges
logarithmically as $T\rightarrow 0$. This pathology is typical of spherical
models, even in the short-ranged uniform case.

Now we analyse other thermodynamic quantities in order to characterize the
system and for comparison with the Ising case \cite{SNA,CL}. The density $d=\frac 12\left[
\left\langle s_{1i}s_{2i}\right\rangle +1\right] _J$ is given by $-\frac{%
\partial f}{\partial \mu }$: 
\begin{equation}
d=\left\{ 
\begin{array}{ll}
1-\frac 1{z_s+2\beta \mu } & P \\ 
1-\frac 1{2\left( \beta +\beta \mu \right) } & SG
\end{array}
\right.  \label{dens}
\end{equation}
it is represented in Fig. 2 as a function of $T$ for several values of $\mu $%
. In the large temperature limit $d$ approaches the value $1/2$ for each $\mu $. For $T=0$ we get $d=q=\vartheta \left( \mu +1\right) $ with $d=q=0$ for $%
\mu =-1$; note that, unlike the Ising version, there is no interval of $\mu $
values where $0<d<1$. In the spin glass phase a partial freezing takes place
($d<q<1$), except at zero temperature where the system is fully frozen ($%
d=q=1$). For $\mu \rightarrow \infty $ the model approaches the spherical SK
limit \cite{KTJ}: $T_c=1$, $d=1$, $q=1-T$.
\begin{figure}[h]
\begin{center}
\epsfxsize=7cm
\epsffile{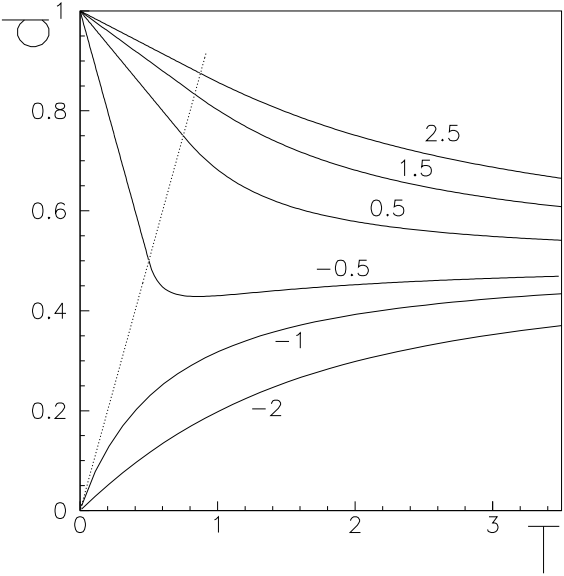}
\caption{The density $d$ vs. $T$ for several values of $\mu $. The intersection of the curves with the dotted line locates the critical temperature $T_{c}\left( \mu \right) $. Thus, the initial straight part of the curves for $\mu >-1$ corresponds to temperatures for which the system is in the glassy phase (bottom of Eq. (\ref{dens})). In the large $T$ limit $d$ goes to the value $1/2$ for each $\mu $. Note also that at zero temperature $d=1$ for $\mu >-1$ and $d=0$ for $\mu \leq -1$.}  
\end{center}
\end{figure}
The compressibility $k=\frac 1\beta \frac{\partial d}{\partial \mu }$ is
found to be: 
\begin{equation}
k=\left\{ 
\begin{array}{ll}
\frac{2\left( 1-\frac 1{z_s+2\beta \mu }\right) }{\left( z_s+2\beta \mu
\right) ^2\left( 1-\frac 1{z_s+2\beta \mu }\right) +\sqrt{z_s^2-4\beta ^2}}
& P \\ 
\frac 1{2\left( \beta +\beta \mu \right) ^2} & SG
\end{array}
\right.  \label{comp}
\end{equation}
its plot as a function of $T$ is given in Fig. 3. It's evident a cusp at the critical temperature $T_{c}\left( \mu \right) $ whose  height diverges in the limit $\mu
\rightarrow -1^{+}$. $k$ goes to the value $1/4$ for each $\mu $ in the large temperature limit. For zero temperature $k=0$ for each $\mu $.
\begin{figure}[h]
\begin{center}
\epsfxsize=7cm
\epsffile{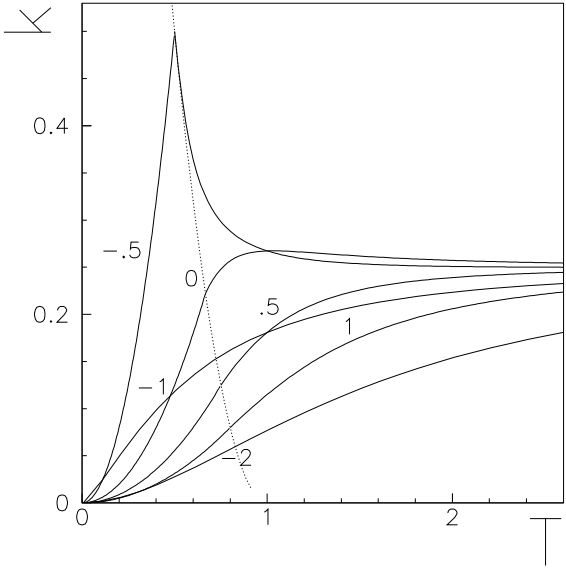}
\caption{The compressibility $k$ vs. $T$ for several values of $\mu $. A cusp is evident at the critical temperature $T_c\left( \mu \right) $ at least for values of $\mu $ close to $-1$. The height of the cusp diverges as $\mu
\rightarrow -1^{+}$. The intersection of the dotted line with the curves for $\mu >-1$ locates the critical temperature $T_{c}\left( \mu \right) $; $k$ increases with the square of $T$ in the glassy phase (bottom of Eq. (\ref{comp})). For $T\rightarrow \infty$ $k$ goes to $1/4$ for each $\mu $.}
\end{center}
\end{figure}
Finally the specific heat $c$ is: 
\begin{equation}
c=\left\{ 
\begin{array}{ll}
1-\frac{z_s}2-\frac{\sqrt{z_s^2-4\beta ^2}\left[ \left( z_s+2\beta \mu
\right) ^2\left( 1-\frac 1{z_s+2\beta \mu }\right) +\beta \mu \right]
-\left( z_s+2\beta \mu \right) ^2}{\left( z_s+2\beta \mu \right) ^2\left( 1-%
\frac 1{z_s+2\beta \mu }\right) +\sqrt{z_s^2-4\beta ^2}} & P \\ 
1 & SG
\end{array}
\right.  \label{spec}
\end{equation}
it presents a cusp at the transition for $\mu >-1$, while for $T=0$ is
discontinuous since $c=\frac 12$ for $\mu \leq -1$ (Fig. 4). The large $\mu $
limit is given by $c=\frac 12+\frac 1{2T^2}$ for $T>1$.
\begin{figure}[h]
\begin{center}
\epsfxsize=7cm
\epsffile{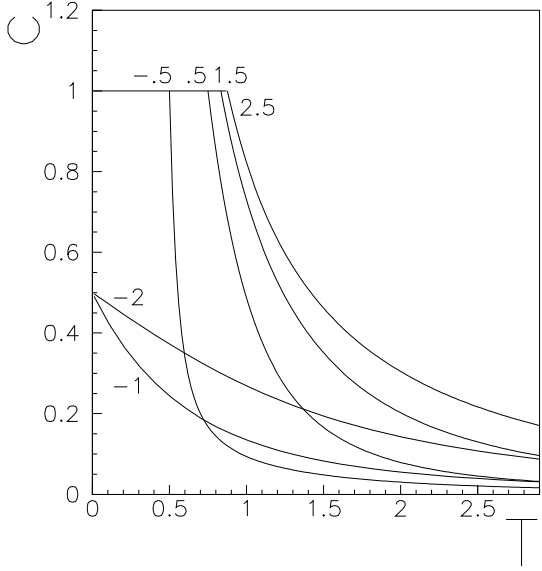}
\caption{The specific heat $c$ vs. $T$ for several values of $\mu $. Note the cusp at the transition temperature $T_c\left( \mu \right) $ for $\mu >-1$. $c=1$ in the glassy phase (bottom of Eq. (\ref{spec})). In the zero $T$ limit $c=1/2$ for each $\mu \leq -1$.}
\end{center}
\end{figure}

\subsection{In the presence of a magnetic field}

Adding to the hamiltonian (\ref{BEGs ham}) a magnetic field term, $-\sum_i%
\frac{h_i}2\left( s_{1i}+s_{2i}\right) $, the saddle point equation is
modified by adding $\int d\lambda \rho \left( \lambda \right) \frac{\left(
\beta h_\lambda \right) ^2}{\left( z-\beta \lambda \right) ^2}$ to the left
hand side of Eq. (\ref{saddle-eq}). Assuming a uniform field, $h_i=h$, one
can replace $h_\lambda ^2$ by its average value $h^2$. Thus one finds that
for $h\neq 0$ there is no transition, since in this case $z_s$ never reaches
the branch point $2\beta $.

Let us now compute the diluted susceptibilitise in zero field. The linear
one obeys the diluted Curie law in the disordered phase and is constant for $%
T<T_c\left( \mu \right) $, so to display a cusp crossing the critical line
(Fig. 5): 
\begin{equation}
\chi =\left. -\frac{\partial ^2f}{\partial h^2}\right| _{h=0}=\left\{ 
\begin{array}{ll}
\beta \left( 1-\frac 1{z_s+2\beta \mu }\right) & P \\ 
1 & SG
\end{array}
\right.  \label{susc}
\end{equation}
notice that the previous result can be obtained also from the linear
response theorem $\chi =\beta \left( d-q\right) $. The zero temperature
expression is $\chi =-\mu -\sqrt{\mu ^2-1}$ for $\mu \leq -1$.
\begin{figure}[h]
\begin{center}
\epsfxsize=7cm
\epsffile{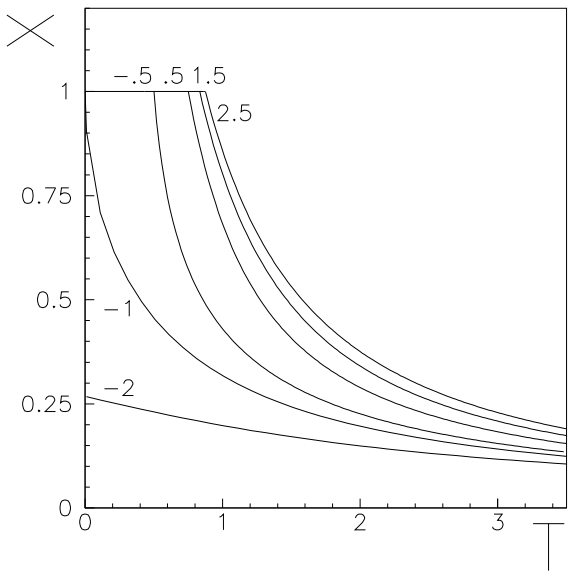}
\caption{The susceptibility $\chi $ vs. $T$ for several values of $\mu $. Note the cusp at the transition temperature $T_c\left( \mu \right) $ for $\mu >-1$. $\chi =1$ in the glassy phase (bottom of Eq. (\ref{susc})). In the zero $T$ limit $\chi =-\mu -\sqrt{\mu ^2-1}$ for $\mu \leq -1$.} 
\end{center}
\end{figure}
The spin glass susceptibility is given by: 
\begin{equation}
\chi _{SG}=\sum_{\lambda \mu }\left( \frac{\partial ^2f}{\partial h_\lambda
\partial h_\mu }\right) _{h=0}^2=\left\{ 
\begin{array}{ll}
\frac 12\left( \frac{z_s}{\sqrt{z_s^2-4\beta ^2}}-1\right) & P \\ 
\infty & SG
\end{array}
\right.
\end{equation}
coming from high temperatures it diverges at the critical line as $\frac 1{%
T-T_c\left( \mu \right) }$ and remains infinite in the whole frozen phase.
For zero temperature it is given by $\chi _{SG}=\frac 12\left( -\frac \mu {%
\sqrt{\mu ^2-1}}-1\right) $ for $\mu \leq -1$. The large $\mu $ limit is $%
\chi _{SG}=\frac 1{T^2-1}$ for $T>1$.

\section{Langevin dynamics}

Now we deal with the Langevin relaxional dynamics of the model. Let us
assume that the two spin fields evolve via usual Langevin equations, which
when projected onto the basis $\lambda $ become: 
\begin{equation}
\left\{ 
\begin{array}{c}
\frac{ds_{1\lambda }}{dt}=\left( \frac \lambda 4-\frac{z_1\left( t\right) }2%
\right) s_{1\lambda }+\left( \frac \lambda 4+\frac \mu 2\right) s_{2\lambda
}+h_{1\lambda }\left( t\right) +\xi _{1\lambda }\left( t\right) \\ 
\frac{ds_{2\lambda }}{dt}=\left( \frac \lambda 4+\frac \mu 2\right)
s_{1\lambda }+\left( \frac \lambda 4-\frac{z_2\left( t\right) }2\right)
s_{2\lambda }+h_{2\lambda }\left( t\right) +\xi _{2\lambda }\left( t\right)
\end{array}
\right.  \label{Lan.eqs.}
\end{equation}
where $z_a\left( t\right) \ \ a=1,2$ are two time-dependent Lagrange
multipliers enforcing the spherical constraints, $h_{a\lambda }\left(
t\right) $ two external fields interacting with the field $s_{a\lambda }$
and $\xi _{a\lambda }$ the thermal noises with zero mean and correlations $%
\left\langle \xi _{a\lambda }\left( t\right) \xi _{b\mu }\left( t^{\prime
}\right) \right\rangle =2T\delta _{ab}\delta _{\lambda \mu }\delta \left(
t-t^{\prime }\right) \ \ a,b=1,2$. Hereafter we use $\left\langle \text{ }%
\right\rangle $ to represent the average over the thermal noises.

Let us now introduce the quantities of interest, namely the correlation
functions $C_{ab}\left( t,t^{\prime }\right) $ $\left( a,b=1,2\right) $, the
response functions $G_{ab}\left( t,t^{\prime }\right) $ and the density $%
d\left( t\right) $: 
\begin{eqnarray}
C_{ab}\left( t,t^{\prime }\right) &=&\left[ \frac 1N\sum_i\left\langle
s_{ai}\left( t\right) s_{bi}\left( t^{\prime }\right) \right\rangle \right]
_J=\int d\lambda \,\rho \left( \lambda \right) \left\langle s_{a\lambda
}\left( t\right) s_{b\lambda }\left( t^{\prime }\right) \right\rangle \\
G_{ab}\left( t,t^{\prime }\right) &=&\left[ \frac 1N\sum_i\left. \frac{%
\delta \left\langle s_{ai}\left( t\right) \right\rangle }{\delta
h_{bi}\left( t^{\prime }\right) }\right| _{h=0}\right] _J=\int d\lambda
\,\rho \left( \lambda \right) \left. \frac{\delta \left\langle s_{a\lambda
}\left( t\right) \right\rangle }{\delta h_{b\lambda }\left( t^{\prime
}\right) }\right| _{h=0} \\
d\left( t\right) &=&\frac 12\left( C_{12}\left( t,t\right) +1\right)
\end{eqnarray}

A quite general procedure allows to derive from (\ref{Lan.eqs.}) closed
equations for these functions as saddle point solutions of a dynamical
generating functional \cite{SZ,KT}. Using this procedure one implicitely
assumes the initial lattice fields $s_{ai}\left( 0\right) $ as random
variables with a gaussian distribution of zero mean and variance $\overline{%
s_{ai}\left( 0\right) s_{bi}\left( 0\right) }=1+2\left( 1-\delta
_{ab}\right) \left( d\left( 0\right) -1\right) $ (the overbar stands for the
average over the random initial conditions); the same is valid for $%
s_{a\lambda }\left( 0\right) $, since the rotational invariance of the
gaussian distribution. However, in our case these functional techniques can
be avoided, as much as \cite{CDP,CD}. Due to its linearity, the Langevin
system (\ref{Lan.eqs.}) can be explicitely solved for given noises, thus the
various dynamical quantities can be evaluated averaging over the noises and
the eigenvalues $\lambda $. In the following we choose the initial
conditions indicated previously. The dynamical model can be solved for any $d\left( 0\right) $; but it can be shown that the value of $d\left( 0\right) $ does not influence the long time behaviour for nonzero temperatures, so we'll take for simplicity $d\left( 0\right) =1$.

For zero external fields the dynamical model is for each time symmetric
under the exchange of the two spin fields. Since below we'll discuss the
dynamics only in this case, we can exploit this symmetry and limit ourselves
to solve Eqs. (\ref{Lan.eqs.}) for $z_1\left( t\right) =z_2\left( t\right) $%
. The solution is then given by: 
\begin{eqnarray}
s_{a\lambda }\left( t\right) &=&\frac 1{2\sqrt{\Gamma \left( t\right) }}%
\left\{ e^{\lambda t/2}\left[ s_{1\lambda }\left( 0\right) +s_{2\lambda
}\left( 0\right) \right] +\eta _ae^{-\mu t}\left[ s_{1\lambda }\left(
0\right) -s_{2\lambda }\left( 0\right) \right] \right.  \nonumber \\
&&+\int_0^tdu\,e^{\lambda \left( t-u\right) /2}\sqrt{\Gamma \left( u\right) }%
\left[ \xi _{1\lambda }\left( u\right) +h_{1\lambda }\left( u\right) +\xi
_{2\lambda }\left( u\right) +h_{2\lambda }\left( u\right) \right]  \nonumber
\\
&&\left. +\eta _a\int_0^tdu\,e^{-\mu \left( t-u\right) }\sqrt{\Gamma \left(
u\right) }\left[ \xi _{1\lambda }\left( u\right) +h_{1\lambda }\left(
u\right) -\xi _{2\lambda }\left( u\right) -h_{2\lambda }\left( u\right)
\right] \right\}  \label{sol of lan eq.}
\end{eqnarray}
where $\eta _a=\delta _{a,1}-\delta _{a,2}$ and 
\begin{equation}
\Gamma \left( t\right) =\exp \int_0^tz\left( u\right) du\qquad z\left(
t\right) =z_1\left( t\right) -\mu =z_2\left( t\right) -\mu
\end{equation}

As a consequence of the above symmetry, in the absence of external fields
the four correlation functions coincide mutually, so we have to study only
two of them: $C_{1a}\;\left( a=1,2\right) $. The same occurs for the
relative response functions. Notice that, when written in the original
variables $s_i$ and $n_i$, $C_{11}\left( t,t^{\prime }\right) $ is just the
spin-spin correlation function, while $C_{12}\left( t,t^{\prime }\right) $
is a rather strange correlator made by a combination of spin and density
variables. Instead, the density-density connected correlation function,
which enters in the schematic version of MCT \cite{MCT}, involves 4-spin
functions in the formalism of the lattice fields $s_1,s_2$:

\begin{equation}
C_{nn}\left( t,t^{\prime }\right) =\left[ \frac 1{4N}\sum_i\left[
\left\langle s_{1i}\left( t\right) s_{2i}\left( t\right) s_{1i}\left(
t^{\prime }\right) s_{2i}\left( t^{\prime }\right) \right\rangle
-\left\langle s_{1i}\left( t\right) s_{2i}\left( t\right) \right\rangle
\left\langle s_{1i}\left( t^{\prime }\right) s_{2i}\left( t^{\prime }\right)
\right\rangle \right] \,\right] _J
\end{equation}
However, since the model is quadratic, this quantity is readily related to $%
C_{1a}$ and we get $C_{nn}\left( t,t^{\prime }\right) =\frac 14\left[
C_{11}\left( t,t^{\prime }\right) ^2+C_{12}\left( t,t^{\prime }\right)
^2\right] $.

From Eq. (\ref{sol of lan eq.}), taking into account of the previous initial
conditions, we find for $t\geq t^{\prime }$ and zero external fields: 
\begin{mathletters}
\begin{eqnarray}
C_{1a}\left( t,t^{\prime }\right)  &=&\frac 2{\sqrt{\Gamma \left( t\right)
\Gamma \left( t^{\prime }\right) }}\left[ \frac{I_1\left( t+t^{\prime
}\right) }{t+t^{\prime }}+T\int_0^{t^{\prime }}du\left( \frac{I_1\left(
t+t^{\prime }-2u\right) }{t+t^{\prime }-2u}+\frac{\eta _ae^{-\mu \left(
t+t^{\prime }-2u\right) }}2\right) \Gamma \left( u\right) \right] 
\label{c1a} \\
G_{1a}\left( t,t^{\prime }\right)  &=&\sqrt{\frac{\Gamma \left( t^{\prime
}\right) }{\Gamma \left( t\right) }}\left[ \frac{I_1\left( t-t^{\prime
}\right) }{t-t^{\prime }}+\frac{\eta _a}2e^{-\mu \left( t-t^{\prime }\right)
}\right]   \label{g1a}
\end{eqnarray}
In these formulas $I_1$ is the modified Bessel function of order $1$;
we have used that $\int d\lambda \rho \left( \lambda \right) e^{\lambda t}=%
\frac{I_1\left( 2t\right) }t$. Instead, the function $\Gamma \left( t\right) 
$ is still indeterminate; it can be computed self-consistently, as solution
of the integral equation obtained by enforcing the spherical constraint $%
C_{11}\left( t,t\right) =1$: 
\end{mathletters}
\begin{equation}
\Gamma \left( t\right) =\frac{I_1\left( 2t\right) }t+T\int_0^tdu\left[ \frac{%
I_1\left( 2\left( t-u\right) \right) }{t-u}+e^{-2\mu \left( t-u\right)
}\right] \Gamma \left( u\right)   \label{ga eq.}
\end{equation}
Taking into account of Eq. (\ref{ga eq.}), one can get the following useful
expressions for $C_{1a}\left( t,t^{\prime }\right) $ and $d\left( t\right) $%
: 
\begin{mathletters}
\begin{eqnarray}
C_{1a}\left( t,t^{\prime }\right)  &=&\frac{\Gamma \left( \frac{t+t^{\prime }%
}2\right) }{\sqrt{\Gamma \left( t\right) \Gamma \left( t^{\prime }\right) }}%
\left\{ 1+2\delta _{a,2}\left[ d\left( \frac{t+t^{\prime }}2\right)
-1\right] \right.   \nonumber \\
&&\left. -2T\int_{t^{\prime }}^{\left( t+t^{\prime }\right) /2}du\left[ 
\frac{I_1\left( t+t^{\prime }-2u\right) }{t+t^{\prime }-2u}+\frac{\eta _a}2%
e^{-\mu \left( t+t^{\prime }-2u\right) }\right] \frac{\Gamma \left(
u\right) }{\Gamma \left( \frac{t+t^{\prime }}2\right) }\right\} 
\label{c1a-bis} \\
d\left( t\right)  &=&1-\frac T{\Gamma \left( t\right) }\int_0^tdue^{-2\mu
\left( t-u\right) }\Gamma \left( u\right)   \label{den(t)}
\end{eqnarray}
We note from these formulas that the behaviour of $C_{1a}\left( t,t^{\prime
}\right) $ at large times can be deduced from that of $\Gamma $ and $d$;
instead $d\left( t\right) $ takes contributions from $\Gamma $ also at low
times. However we will able to compute exactly $\Gamma \left( t\right) $ for
each time and then also $d\left( t\right) $. In the limit $\mu \rightarrow
\infty $ one can neglect the last term in Eqs. (\ref{c1a}),...,(\ref{den(t)}%
); hence $d\left( t\right) =1$, the two correlation functions and the two
response functions coincide, recovering the case of spherical SK model \cite
{CD}.

\section{Computing the function $\Gamma $ and the density}

Firstly, we get $\Gamma \left( t\right) $ solving the integral equation (\ref
{ga eq.}) by Laplace transform techniques. Following \cite{CD}, one could
solve Eq. (\ref{ga eq.}) using a suitable expansion which is valid in the spin
glass phase. However, we proceed by a different technique in order to obtain 
$\Gamma \left( t\right) $ in the whole phase diagram. We find that the
structure of the function $\Gamma \left( t\right) $ is related to the
general roots of Eq. (\ref{saddle-eq}), discussed in all details in App. A.

Taking the Laplace transform of Eq. (\ref{ga eq.}) and using the convolution
theorem, after some algebra we put $\Gamma \left( s\right) $ in the form: 
\end{mathletters}
\begin{equation}
\Gamma \left( s\right) =\beta \left[ \frac{P\left( \beta s\right) }{C\left(
\beta s\right) }-\beta \frac{Q\left( \beta s\right) }{C\left( \beta s\right) 
}\left( \frac{s-\sqrt{s^2-4}}2\right) \right]  \label{ga-lptr}
\end{equation}
where $P\left( z\right) =\left( z+2\beta \mu \right) ^2$, $Q\left( z\right)
=\left( z+2\beta \mu \right) \left( z+2\beta \mu -1\right) $ and $C\left(
z\right) $ is a third degree polynomial given by Eq. (\ref{cubic}). We see
that $\Gamma \left( s\right) $ is written in terms of rational functions and 
$\frac{s-\sqrt{s^2-4}}2$ which is the Laplace transform of $\frac{I_1\left(
2t\right) }t$; thus in this form we can take readily the inverse transform.
If $C\left( z\right) $ has distinct roots $a_k$\ $\left( k=1,2,3\right) $\
this is given by: 
\begin{equation}
\Gamma \left( t\right) =\sum_{k=1}^3\left[ \phi _k^P\left( a_k\right)
e^{\left( a_k/\beta \right) t}-\beta \phi _k^Q\left( a_k\right)
\int_0^te^{\left( a_k/\beta \right) \left( t-u\right) }\frac{I_1\left(
2u\right) }udu\right]  \label{ga ex1}
\end{equation}
where $\phi _k^P\left( a_k\right) =\frac{P\left( a_k\right) }{C^{\prime
}\left( a_k\right) }$ and $\phi _k^Q\left( a_k\right) =\frac{Q\left(
a_k\right) }{C^{\prime }\left( a_k\right) }$ with $k=1,2,3$. The case of
degenere roots (lines $l_i$ in Fig. 8) can be treated analogously and
presents no qualitative differences; so we leave it in App. B.

The integral appearing in Eq. (\ref{ga ex1}) can be manipulated using the
integral representation of the modified Bessel function $I_1$; we obtain 
\begin{equation}
\int_0^te^{-c\omega }\frac{I_1\left( \omega \right) }\omega d\omega =c-\sqrt{%
c^2-1}-e^{-\left( c-1\right) t}J_c\left( t\right) ,\qquad J_c\left( t\right)
=\frac 1\pi \int_{-1}^1dx\frac{\sqrt{1-x^2}}{c-x}e^{\left( x-1\right) t}
\label{I(t)}
\end{equation}
where $c$ is in general complex but $\notin \left] -1,1\right[ $. For long
time the integral $J_c\left( t\right) $ can be evaluated analytically by a suitable expansion. For $%
t\gg \frac 1{\left| c-1\right| }$ we get: 
\begin{equation}
J_c\left( t\right) \simeq \frac 1{\sqrt{2\pi t^3}\left( c-1\right) }\left[ 1-%
\frac 3{2t}\left( \frac 1{c-1}+\frac 14\right) \right]  \label{gei1}
\end{equation}
When $c$ is real and close to $1$, a new time regime exists for $1\ll t\ll 
\frac 1{c-1}$. In this regime we find a different behaviour of the integral: 
\begin{equation}
J_c\left( t\right) \simeq \sqrt{\frac 2{\pi t}}\left[ 1-\sqrt{\pi t\left(
c-1\right) }+2t\left( c-1\right) -\frac 1{8t}\right]  \label{gei2}
\end{equation}
In particular, in the limit $c\rightarrow 1^{+}$ this regime holds for each $%
t\gg 1$ and one has $J_c\left( t\right) \simeq \sqrt{\frac 2{\pi t}}$. Note that the leading term in (\ref{gei1}) or (\ref{gei2}) is enough for the
following discussion; we retain the higher order terms in the expansions only for
the numerical calculations of the next section. \newline
Let us come back to $\Gamma \left( t\right) $. Taking into account of Eq. (%
\ref{I(t)}), it can be rewritten as: 
\begin{equation}
\Gamma \left( t\right) =\sum_{k=1}^3\left[ S\left( a_k\right) e^{\left(
a_k/\beta \right) t}+\phi _k^Q\left( a_k\right) \beta e^{2t}J_{a_k/2\beta
}\left( 2t\right) \right]  \label{ga fin}
\end{equation}
where: 
\begin{eqnarray}
S\left( a_k\right) &=&\phi _k^P\left( a_k\right) -\phi _k^Q\left( a_k\right)
\left( \frac{a_k-\sqrt{a_k^2-4\beta ^2}}2\right)  \nonumber \\
&=&\left\{ 
\begin{array}{ll}
\frac{\left( a_k+2\beta \mu \right) ^2-\beta ^2\left( a_k+2\beta \mu
-1\right) ^2}{C^{\prime }\left( a_k\right) }\quad & \text{if }a_k\text{
satisfies Eq. (\ref{saddle-eq})} \\ 
\frac{C\left( a_k\right) }{C^{\prime }\left( a_k\right) }=0 & \text{else}
\end{array}
\right.  \label{deS}
\end{eqnarray}
Once $\Gamma \left( t\right) $ is known, the function $d\left( t\right) $
can be also exactly evaluated. Replacing Eq. (\ref{ga fin}) in (\ref{den(t)}) we
find: 
\begin{eqnarray}
d\left( t\right) &=&1-\frac T{2\Gamma \left( t\right) }\sum_{k=1}^3\left\{ 
\frac{S\left( a_k\right) }{a_k/2\beta +\mu }\left( e^{\left( a_k/\beta
\right) t}-e^{-2\mu t}\right) \right.  \nonumber \\
&&\left. +\frac{\phi _k^Q\left( a_k\right) \beta }{a_k/2\beta +\mu }\left[
e^{2t}\left( J_{a_k/2\beta }\left( 2t\right) -J_{-\mu }\left( 2t\right)
\right) -e^{-2\mu t}\left( \frac{a_k-\sqrt{a_k^2-4\beta ^2}}{2\beta }+\mu +%
\sqrt{\mu ^2-1}\right) \right] \right\}  \label{den-cal fin}
\end{eqnarray}

From Eq. (\ref{deS}) we see that the exponentials with roots $a_k$ not
satisfying Eq. (\ref{saddle-eq}) play no role. To know what and how many
exponentials make up $\Gamma \left( t\right) $ and $d\left( t\right) $ in
the different regions of the phase diagram, one has just to consult the
table (\ref{table}) and Fig. 8. Computing the large time limit of $\Gamma
\left( t\right) $ and $d\left( t\right) $, we find only a few different
behaviours which we list now. For zero temperature we have $\Gamma \left(
t\right) =\frac{I_1\left( 2t\right) }t$ and $d\left( t\right) =1$ for each $%
t $ and $\mu $. In the spin glass phase and for $T=T_c\left( \mu \right) $
we retain in (\ref{ga fin},\ref{den-cal fin}) only the dominant exponential $%
e^{2t}$. Evaluating $J_{a_k/2\beta }\left( 2t\right) $ and $J_{-\mu }\left(
2t\right) $ by the leading term in (\ref{gei1}), and then using the identity
(\ref{rel.roots}), we get after some algebra: 
\begin{mathletters}
\begin{eqnarray}
\Gamma \left( t\right) &\simeq &\frac{e^{2t}}{\sqrt{4\pi t^3}}\sum_{k=1}^3%
\frac{\beta ^2\phi _k^Q\left( a_k\right) }{a_k-2\beta }=\frac{e^{2t}}{\sqrt{%
4\pi t^3}}\frac d{q_{EA}^2}  \label{ga-tltc} \\
d\left( t\right) &\simeq &1-\frac T{2\Gamma \left( t\right) }\sum_{k=1}^3%
\frac{\beta ^2\phi _k^Q\left( a_k\right) }{a_k+2\beta \mu }\left[ \frac{%
e^{2t}}{\sqrt{4\pi t^3}}\left( \frac{2\beta }{a_k-2\beta }+\frac 1{\mu +1}%
\right) \right] =d
\end{eqnarray}
where $q_{EA}$ is the Edwards-Anderson parameter coinciding with (\ref{q})
and $d$ the equilibrium density given by the bottom of (\ref{dens}). At the
critical transition line, only a slight difference occurs with respect to
the previous case: the integral $J_{a_1/2\beta }\left( 2t\right) =J_1\left(
2t\right) $, given by (\ref{gei2}), prevails over the other ones for large $%
t $. Then using that $\phi _1^Q\left( 2\beta \right) =1$ one has $\Gamma
\left( t\right) \simeq \beta \frac{e^{2t}}{\sqrt{\pi t}}$. Finally, in the
nonglassy phase we get readily: $\Gamma \left( t\right) \simeq S\left(
z_s\right) e^{\left( z_s/\beta \right) t}$, where $z_s$ is the equilibrium
solution of Eq. (\ref{saddle-eq}). Correspondently, the density tends to its
equilibrium value given by the top of Eq. (\ref{dens}).

\section{Dynamics in the nonglassy phase}

Now we specialize the general expressions of the dynamical quantities
obtained previously to the case of the nonglassy phase. It is convenient to
put in evidence the large time dominant exponential $e^{\left( z_s/\beta
\right) t}$ in Eq. (\ref{ga fin}), $\Gamma \left( t\right) =e^{\left(
z_s/\beta \right) t}\Omega \left( t\right) ,$ 
\end{mathletters}
\begin{equation}
\Omega \left( t\right) =S\left( z_s\right) +e^{-2t/\tau }\sum_{k=1}^3\beta
\phi _k^Q\left( a_k\right) J_{a_k/2\beta }\left( 2t\right) +e^{-2t/\tau
_2}S\left( a_2\right) +e^{-2t/\tau _3}S\left( a_3\right)  \label{omega}
\end{equation}
and then replace $\Gamma \left( t\right) $ in Eqs. (\ref{c1a-bis},b),(\ref{g1a}).
We get the following exact expressions: 
\begin{mathletters}
\begin{eqnarray}
C_{1a}\left( t,t^{\prime }\right) &=&\frac{\Omega \left( \frac{t+t^{\prime }}%
2\right) }{\sqrt{\Omega \left( t\right) \Omega \left( t^{\prime }\right) }}%
\left\{ 1+2\delta _{a,2}\left[ d\left( \frac{t+t^{\prime }}2\right)
-1\right] -\frac 1\beta \int_0^{t-t^{\prime }}d\omega \left[ \frac{I_1\left(
\omega \right) }\omega e^{-\omega \left( 1+1/\tau \right) }+\frac{\eta _a}2%
e^{-\omega /\tau ^{\prime }}\right] \frac{\Omega \left( \frac{t+t^{\prime
}-\omega }2\right) }{\Omega \left( \frac{t+t^{\prime }}2\right) }\right\}
\label{c1a-tgtc} \\
G_{1a}\left( t,t^{\prime }\right) &=&\sqrt{\frac{\Omega \left( t^{\prime
}\right) }{\Omega \left( t\right) }}\left[ \frac{I_1\left( t-t^{\prime
}\right) }{t-t^{\prime }}e^{-\left( t-t^{\prime }\right) \left( 1+1/\tau
\right) }+\frac{\eta _a}2e^{-\left( t-t^{\prime }\right) /\tau ^{\prime
}}\right]  \label{g1a-tgtc} \\
d\left( t\right) &=&1-\frac T{2\Omega \left( t\right) }\left\{ \frac{S\left(
z_s\right) }{z_s/2\beta +\mu }\left( 1-e^{-2t/\tau ^{\prime }}\right) +\frac{%
S\left( a_2\right) }{a_2/2\beta +\mu }\left( e^{-2t/\tau _2}-e^{-2t/\tau
^{\prime }}\right) +\frac{S\left( a_3\right) }{a_3/2\beta +\mu }\left(
e^{-2t/\tau _3}-e^{-2t/\tau ^{\prime }}\right) \right.  \nonumber \\
&&\left. +\sum_{k=1}^3\frac{\phi _k^Q\left( a_k\right) \beta }{a_k/2\beta
+\mu }\left[ e^{-2t/\tau }\left( J_{a_k/2\beta }\left( 2t\right) -J_{-\mu
}\left( 2t\right) \right) -e^{-2t/\tau ^{\prime }}\left( \frac{a_k-\sqrt{%
a_k^2-4\beta ^2}}{2\beta }+\mu +\sqrt{\mu ^2-1}\right) \right] \right\}
\label{dens-tgtc}
\end{eqnarray}
In the previous formulas we have introduced the characteristic times: 
\end{mathletters}
\begin{equation}
\tau =\frac 1{z_s/2\beta -1}\quad \tau ^{\prime }=\frac 1{z_s/2\beta +\mu }%
\quad \tau _2=\frac 1{z_s/2\beta -a_2/2\beta }\quad \tau _3=\frac 1{%
z_s/2\beta -a_3/2\beta }  \label{tau}
\end{equation}
We recall that the exponentials with the characteristic times $\tau _2,\tau
_3$ are absent in the regions of the phase diagram where the relative roots $%
a_2,a_3$ does not satisfy Eq. (\ref{saddle-eq}). In this regard see the
table (\ref{table}) and Fig. 8.

\subsection{Equilibrium dynamics}

From the discussion done in Sec. III we deduce that $\tau _2$ and $\tau _3$,
if present, are in any case lower than the largest between $\tau $ and $\tau
^{\prime }$. This implies that the largest of the characteristic times (\ref
{tau}) is given by $\tau _{eq}=$ $\max \left\{ \tau ,\tau ^{\prime }\right\} 
$. In particular, one has $\tau _{eq}=$ $\tau $ or $\tau ^{\prime }$
according to $\mu >-1$ or $<-1$, while for $\mu =-1$: $\tau _{eq}=\tau =\tau
^{\prime }$. The time $\tau _{eq}$ can be identified as the characteristic
equilibration time of the system. In fact for waiting times $t^{\prime
}>\tau _{eq}$ the density (\ref{dens-tgtc}) is practically constant at the
equilibrium value $d$ given by the top of (\ref{dens}), while the two-time
functions (\ref{c1a-tgtc},b) obey TTI and FDT $\left( TG_{1a}\left(
t-t^{\prime }\right) =\frac{\partial C_{1a}\left( t-t^{\prime }\right) }{%
\partial t^{\prime }}\right) $, being given by: 
\begin{mathletters}
\begin{eqnarray}
C_{1a}\left( t-t^{\prime }\right)  &=&1+2\delta _{a,2}\left( d-1\right) -%
\frac 1\beta \int_0^{t-t^{\prime }}d\omega \left[ \frac{I_1\left( \omega
\right) }\omega e^{-\omega \left( 1+1/\tau \right) }+\frac{\eta _a}2%
e^{-\omega /\tau ^{\prime }}\right]   \nonumber \\
&=&e^{-\left( t-t^{\prime }\right) /\tau }\frac{J_{1+1/\tau }\left(
t-t^{\prime }\right) }\beta +\frac{\eta _a\tau ^{\prime }}{2\beta }%
e^{-\left( t-t^{\prime }\right) /\tau ^{\prime }}  \label{c1a-tti} \\
G_{1a}\left( t-t^{\prime }\right)  &=&\frac{I_1\left( t-t^{\prime }\right) }{%
t-t^{\prime }}e^{-\left( t-t^{\prime }\right) \left( 1+1/\tau \right) }+%
\frac{\eta _a}2e^{-\left( t-t^{\prime }\right) /\tau ^{\prime }}
\label{g1a-tti}
\end{eqnarray}
To obtain the second line we have used Eq. (\ref{I(t)}). Note that the
equilibrium correlation functions $C_{1a}\left( t-t^{\prime }\right) $ decay
exponentially to zero with characteristic relaxation time given just by $%
\tau _{eq}$. For $\mu >-1$, $\tau _{eq}=\tau $ diverges at the critical
line, as $\tau \simeq \frac{2T_c\left( \mu \right) ^4}{\left( T-T_c\left(
\mu \right) \right) ^2}$, signaling critical slowing down; this implies that
for each $\mu >-1$ the dynamical transition temperature coincides with the
static one, as occurs in other continuous models \cite{CK1,CD,CLD}. Actually, the integral $J_{1+1/\tau }\left(
t-t^{\prime }\right) $, given by Eq. (\ref{gei1}) or (\ref{gei2}), provides a power law correction to the exponential decay; it goes as $\left( t-t^{\prime }\right) ^{-3/2}$ for $t-t^{\prime }\gg \tau $ or as $\left( t-t^{\prime }\right) ^{-1/2}$ in the critical regime for $t-t^{\prime }\ll \tau $. Finally, $\tau _{eq}$ diverges also for $T\rightarrow 0$ and $\mu \leq -1
$ as $\tau _{eq}=\tau ^{\prime }\simeq 2\beta $.

\subsection{Nonequilibrium dynamics}

For waiting times $t^{\prime }$ lower than $\tau _{eq}$ the system is not
yet in equilibrium and dynamics is described by Eqs. (\ref{c1a-tgtc},b,c);
these equations are formally analogous to those valid in the spin glass
phase that we discuss below. When $\tau _{eq}$ is small, as usually occurs
in the nonglassy phase, the equilibration is fast and the time range $%
t^{\prime }<\tau _{eq}$ represents just a short initial transient. However, $%
\tau _{eq}$ can be made arbitrarily large as soon as one approaches the
critical line, or for very low temperatures and $\mu \leq -1$; in such case
a true nonequilibrium regime appears, although the system is in the nonglassy
phase.

A quite useful way to characterize the relaxation process is by plotting the
integrated response $\chi _{1a}\left( t,t^{\prime }\right) =\int_{t^{\prime
}}^tG_{1a}\left( t,u\right) du$, multiplied by the temperature $T$, 
\end{mathletters}
\begin{equation}
T\chi _{1a}\left( t,t^{\prime }\right) =\frac 1\beta \int_0^{t-t^{\prime
}}d\omega \left[ \frac{I_1\left( \omega \right) }\omega e^{-\omega \left(
1+1/\tau \right) }+\frac{\eta _a}2e^{-\omega /\tau ^{\prime }}\right] \sqrt{%
\frac{\Omega \left( t-\omega \right) }{\Omega \left( t\right) }}
\label{chi1a-tgtc}
\end{equation}
as a function of the correspondent correlation function $C_{1a}\left(
t,t^{\prime }\right) $, given by Eq. (\ref{c1a-tgtc}), for different values of $t^{\prime }$%
. At equlibrium FDT implies a linear shape of these curves according to the
relation: $T\chi _{1a}=1+2\delta _{a,2}\left( d-1\right) -C_{1a}$, while
this is no more true for $t^{\prime }<\tau _{eq}$. We can thus analyse the
changeover from the equilibrium to the nonequilibrium regime by monitoring
how the curves deviate from this straight line. We find two different
behaviours of the system along the critical line. In order to describe them,
for semplicity we focus on the mode $a=1$.

For $T$ close to $T_c\left( \mu \right) $ with $\mu $ not so close to the
value $-1$ (see Fig. 1), one can readily show that for $t$ large but
sufficiently lower than $\tau _{eq}$ the function $\Omega \left( t\right) $
follows its critical behaviour; one has indeed 
\[
\Omega \left( t\right) \simeq \beta \left( \sqrt{\frac 8\tau }+e^{-2t/\tau
}J_{1+1/\tau }\left( 2t\right) \right) 
\]
so that $\Omega \left( t\right) \simeq \frac \beta {\sqrt{\pi t}}$ for $t\ll 
\frac \tau {8\pi }$, which is just the critical behaviour. The time dependence of $\Omega $ would allow to have aging in the correlation function. However, in this case the
plot $T\chi _{11}$ vs. $C_{11}$ does not give much evidence of the
nonequilibrium regime and we do not report it. In fact, for each $\mu >-1$
and finite temperature, even if $t,t^{\prime }<\tau _{eq}$, FDT still holds
when $t-t^{\prime }\ll t^{\prime }$: the function $C_{11}\left( t,t^{\prime
}\right) $ follows its equilibrium expression (\ref{c1a-tti}) and the plot
starting out at the point $\left( C_{11}=1,T\chi _{11}=0\right) $ is linear
again. If $t^{\prime }\gg \tau ^{\prime }\simeq \frac 1{1+\mu }$, one has $%
C_{11}\left( t-t^{\prime }\right) \simeq \frac 1\beta \sqrt{\frac 2{\pi
\left( t-t^{\prime }\right) }}$ so that this regime extends up to small
values of the correlation; thus small deviations from the straight line
occur only in the last part of the plot until the endpoint $\left(
C_{11}=0,T\chi _{11}=1\right) $ is asymptotically reached.

Instead, for low temperatures and $\mu $ near $-1$, the range of $C_{11}$
values in the FDT regime ($t-t^{\prime }\ll t^{\prime }$) decreases and for
zero temperature it disappears being $\chi _{11}=0$ for each $t,t^{\prime }$%
. One finds for $\mu =-1$ and low $T$
\[
\Omega \left( t\right) \simeq \frac 1\beta +e^{-2t/\tau }\frac 1{%
\sqrt{4\pi t^3}} 
\]
with $\tau =2\beta $; thus for $t\ll \left( \frac{%
\tau ^2}{16\pi }\right) ^{1/3}$ one recovers the zero temperature behaviour $%
\Omega \left( t\right) \simeq \frac 1{\sqrt{4\pi t^3}}.$ It may be also
shown that the same behaviour occurs for $\mu <-1$ when $t\ll \frac{\ln
\beta }{2\left( -\mu -1\right) }$. In Fig. 6 we show the curves $T\chi _{11}$
vs. $C_{11}$ for various $t^{\prime }/\tau _{eq}$ obtained in the case $%
\mu =-1$ and low $T$ so that $\tau _{eq}=2\,10^4$. The model exhibits in this case the pattern of interrupted aging. If $T$ is nonzero, i.e. $\tau _{eq}=2\beta $ is finite, the large $t$ limit
of $C_{11}\left( t,t^{\prime }\right) $ and $T\chi _{11}$ is given
respectively by $0$ and $1$. Therefore all the curves starting out at the
point $\left( C_{11}=1,T\chi _{11}=0\right) $, must end up in the same point 
$\left( C_{11}=0,T\chi _{11}=1\right) $. The dependence on $t^{\prime }/\tau
_{eq}$ enters on how the initial and final point are joined. If $t^{\prime
}/\tau _{eq}>1$ a linear plot is obtained, as already said. If $t^{\prime
}/\tau _{eq}<1,$ then the plot covers a very short part of the straight FDT
line and then falls below this line. In particular, if the condition $%
t^{\prime }/\tau _{eq},t/\tau _{eq}\ll 1$ is realized, $C_{11}$ obeys
approximatively the zero temperature scaling form $2^{3/2}\left( t^{\prime }/t\right) ^{3/4}$ and $T\chi _{11}$ is almost
vanishing. The plot follows this shape for a range which is larger the smaller is the value of $t^{\prime }/\tau _{eq}$. Looking at these pieces of the curves only, one could wrongly
conclude that the system is in glassy phase. But, as $t$ becomes larger than 
$\tau _{eq}$ the plot must raise again in order to reach the point $\left(
C_{11}=0,T\chi _{11}=1\right) $; thus the final upword bending of the curves
is a consequence of a finite equilibration time and corresponds to
interrupted aging. In the limiting case $\tau _{eq}=\infty $ aging holds for
all time and the plot obeys $T\chi _{11}=0$ over the entire range of $C_{11}$
values. The behaviour now described is similar to that recently found in the
one-dimensional Ising model with Glauber dynamics \cite{LZ}, but in that
case a less trivial shape of $T\chi \left( C\right) $ is obtained.

Finally, moving along
or slightly above the critical line with $\mu \simeq -1$ for a fixed $t^{\prime }=50$, one obtains the curves shown in the \emph{inset} of Fig. 6. We note that their global shape is unchanged, but now the initial straight part correspondent to the FDT regime gets longer the larger $\mu $.  
\begin{figure}[h]
\begin{center}
\epsfxsize=9cm
\epsffile{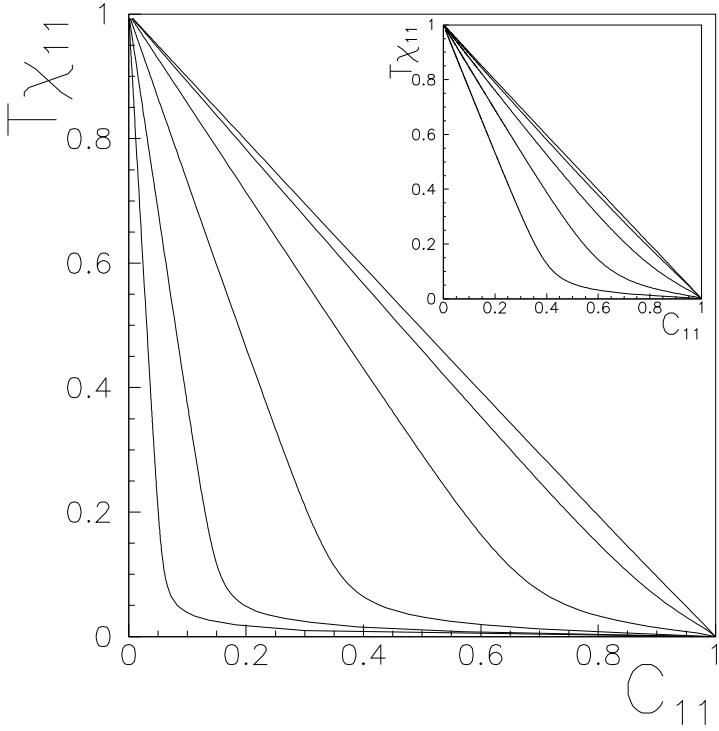}
\caption{$T\chi _{11}$ vs. $C_{11}$ for different values of $t^{\prime
}$, with $\tau _{eq}\simeq 2\beta =2\,10^4$ and $\mu =-1$. We
have taken $t^{\prime }/\tau _{eq}=10^{-x}$ with $x=1,\,1.5,\,2,\,2.5,%
\,3,\,3.5$ for the curves respectively from the right to the left. The plot illustrates the nonlinear dependence of the integrated response $T\chi _{11}$ as a function of $C_{11}$ when $t^{\prime }$ is lower than $\tau _{eq}$. If the condition $%
t^{\prime }/\tau _{eq},t/\tau _{eq}\ll 1$ is realized, $C_{11}$ obeys
approximatively the zero temperature scaling form $2^{3/2}\left( t^{\prime }/t\right) ^{3/4}$ and $T\chi _{11}$ is almost
vanishing, as the system was in the low temperature glassy phase. The plot follows this shape for a range which is larger the smaller is the value of $t^{\prime }/\tau _{eq}$. Then, as a consequence of the finite equilibration time, the plot must raise again in order to reach the point $\left(
C_{11}=0,T\chi _{11}=1\right) $; thus the final upword bending of the curves corresponds to
interrupted aging. The \emph{inset} shows the curves obtained by keeping fixed $t^{\prime } =50$ and varying $\mu $ around $-1$. We have taken $\mu =-1+10^{-x}$ with $x=2, 2.5, 3, 3.5, 4$ for the curves respectively from left to right. The initial straight part of the curves correspondent to the FDT regime increases with $\mu $. } 
\end{center}
\end{figure}
\section{Nonequilibrium dynamics in the glassy phase}

The analysis of the nonequilibrium dynamics in the glassy phase is quite
similar to that of the spherical SK model, discussed in \cite{CD}. As above,
we can put in evidence the dominant exponential in Eq. (\ref{ga fin}), which
now is $e^{2t}$, and then replace $\Gamma \left( t\right) $ in Eqs. (\ref
{c1a-bis},b),(\ref{g1a}). The expressions we get for these quantities can be
obtained by Eqs. (\ref{c1a-tgtc},b,c) for $%
S\left( z_s\right) =0$ and 
\begin{equation}
\tau =\infty \quad \tau ^{\prime }=\frac 1{1+\mu }\quad \tau _2=\frac 1{%
1-a_2/2\beta }\quad \tau _3=\frac 1{1-a_3/2\beta }
\end{equation}
The system is out of equilibrium on each time scale since $\tau _{eq}$ is
infinite. Even if the roots $a_2,a_3$ satisfy Eq. (\ref{saddle-eq}), the
characteristic times $\tau _2$ and $\tau _3$ are very small (lower than $%
\frac 12$) and, in practice, well inside the glassy phase, one has $\Omega
\left( t\right) \simeq \frac 1{\sqrt{4\pi t^3}}\frac d{q_{EA}^2}$ and $%
d\left( t\right) \simeq d$ where $q_{EA}$ is the Edwards-Anderson parameter
coinciding with (\ref{q}) and $d$ the equilibrium density given by the
bottom of (\ref{dens}). About the two-time quantities, for large times we
distinguish the following two regimes:

\begin{itemize}
\item  FDT regime for $t\simeq t^{\prime }$ with $t-t^{\prime }\ll t^{\prime
}$: 
\begin{mathletters}
\begin{eqnarray}
C_{1a}\left( t-t^{\prime }\right)  &=&1+2\delta _{a,2}\left( d-1\right) -%
\frac 1\beta \int_0^{t-t^{\prime }}d\omega \left[ \frac{I_1\left( \omega
\right) }\omega e^{-\omega }+\frac{\eta _ae^{-\omega /\tau ^{\prime }}}2%
\right]   \nonumber \\
&=&q_{EA}+\frac{J_1\left( t-t^{\prime }\right) }\beta +\frac{\eta _a\tau
^{\prime }}{2\beta }e^{-\left( t-t^{\prime }\right) /\tau ^{\prime }}
\label{c1a-tti1} \\
G_{1a}\left( t-t^{\prime }\right)  &=&\frac{I_1\left( t-t^{\prime }\right) }{%
t-t^{\prime }}e^{-\left( t-t^{\prime }\right) }+\frac{\eta _a}2e^{-\left(
t-t^{\prime }\right) /\tau ^{\prime }}  \label{g1a-tti1}
\end{eqnarray}
\end{mathletters}
where $q_{EA}$ is the Edwards-Anderson parameter coinciding with (\ref{q})
and $d$ the equilibrium density given by the bottom of (\ref{dens}). In this
regime the properties TTI and FDT are satisfied. In the large $t-t^{\prime }$
limit the two correlation functions have a power law decay to the value $q_{EA}$ as $%
C_{1a}\left( t-t^{\prime }\right) -q_{EA}\simeq \frac 1\beta \sqrt{\frac 2{%
\pi \left( t-t^{\prime }\right) }}$; the reaching of $q_{EA}$ determines
the end of this regime.
\item  Aging regime for $t>t^{\prime }$, $\lambda =\frac{t^{\prime }}t\simeq
0$: 
\begin{mathletters}
\begin{eqnarray}
C_{1a}\left( t,t^{\prime }\right)  &=&2^{3/2}\frac{\lambda ^{3/4}}{\left(
1+\lambda \right) ^{3/2}}q_{EA}\simeq 2^{3/2}\lambda ^{3/4}q_{EA}
\label{c1a-ag} \\
G_{1a}\left( t,t^{\prime }\right)  &=&\frac 1{\sqrt{2\pi }}\frac{\lambda
^{-3/4}}{\left( 1-\lambda \right) ^{3/2}}t^{-3/2}\simeq \frac 1{\sqrt{2\pi }}%
\lambda ^{-3/4}t^{-3/2}  \label{g1a-ag}
\end{eqnarray}
\end{mathletters}
Here FDT and TTI are violated. The two correlation functions coincide and
have a slow decay to zero obeying power scaling. For spin glass like models,
in this regime FDT can be generalized to $TG\left( t,t^{\prime }\right)
=X\left( C\left( t,t^{\prime }\right) \right) \frac{\partial C\left(
t,t^{\prime }\right) }{\partial t^{\prime }}$, where $X$ is the
fluctuation-dissipation ratio assumed to depend on time arguments only
through $C$ and the function $X\left( C\right) $ is characteristic of the
model \cite{CK,CK1,CD,CLD}. In our case from Eqs. (\ref{c1a-ag},b) we get $%
X_{1a}\left( C_{1a}\right) =0$, as much as \cite{CD}.

\end{itemize}
Fig. 7 shows an example of plot $T\chi _{1a}$ vs. $C_{1a}$ obtained when the
system in this phase. For larger $t^{\prime }$ the curves approach the
asymptots 
\begin{equation}
T\chi _{1a}=\left\{ 
\begin{array}{ll}
1+2\delta _{a,2}\left( d-1\right) -C_{1a}\quad & q_{EA}<C_{1a}<1+2\delta
_{a,2}\left( d-1\right) \\ 
1+2\delta _{a,2}\left( d-1\right) -q_{EA} & 0<C_{1a}<q_{EA}
\end{array}
\right.  \label{asymptotics}
\end{equation}
which can be derived by Eqs. (\ref{c1a-tti1},b) and (\ref{c1a-ag},b); they
correspond to the two regimes discussed previously. For zero temperature the
FDT regime is absent $\left( q_{EA}=1\right) $.
\begin{figure}[h]
\begin{center}
\epsfxsize=7cm
\epsffile{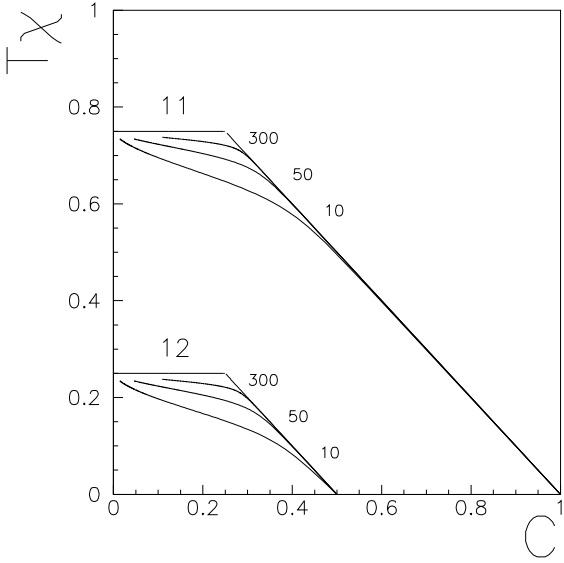}
\caption{$T\chi _{11}$ vs. $C_{11}$ and $T\chi _{12}$ vs. $C_{12}$ for different values of $t^{\prime }$, with $T=.5$ and $\mu =0$. Notice that for
larger $t^{\prime }$ the curve approach the asymptots given by Eq. (\ref
{asymptotics}), corresponding to the two large time regimes described in the text.}
\end{center}
\end{figure}
\section{Conclusions}

We have defined a spherical version of the frustrated BEG model by enforcing
spherical constraints to suitable Ising variables. The main advantage of such a model consists of its relative
simplicity which allows to obtain a full analytical solution of the
equilibrium properties and of Langevin relaxation dynamics at mean field
level. As a first approach to
the model, we have studied in detail the case $K=0$, but the
same kind of analysis can be equivantely carried out in the whole range of
the parameter $K$ \cite{FW}.

Specifically, we have showed that quantities like the density, the
compressibility or the density-density correlator, which are used in the
study of liquids, can be introduced in this framework and exactly evaluated.
This is convenient in the attempt to make a more suitable
description of the glass transition, using the theoretical background
developed for spin glasses. In this regard we note that the technique of
sphericization used here, could be applied and tested in other diluted spin
glass models. 

The equilibrium phase diagram, shown in Fig. 1, is rather simple. The line of
discontinuous transition found in the Ising version of the model \cite{SNA,CL}, in this spherical case with $K=0$ collapses in a single point at $\left( \mu =-1,T=0\right) $.
Then the transition for $\mu >-1$ is always continuous, as occurs in the
spherical SK model \cite{KTJ}.

The study of the Langevin dynamics from a random initial condition has led to the same phase diagram. Neverthless, this study has displayed a very interesting behaviour. In the glassy phase it
essentially reproduces the findings of  \cite{CD} for the sperical SK model.
Our exact analysis of the dynamics in the whole phase diagram has taken into
account of all the characteristic time scales of the system, which become
important in the preasymptotic time regime. We have pointed out that a nonequilibrium
regime, usually associated to the glassy phase, is possible also in the
nonglassy phase for waiting times lower than the characteristic equilibration time. As for the glassy phase, it is characterized by a violation of FDT, manifested by a non linear behaviour of the
integrated response as a function of the correlation. In particular, we have seen that the presence of such nonequilibrium regime becomes more evident in the region near the point $\left(
\mu =-1,T=0\right) $, where the system displays explicitly the pattern of interrupted aging (Fig. 6). From an experimental or numerical point of view, this behaviour could make rather ambigous the onset of the glassy phase if the system is probed on restricted time scales.

\appendix 

\section{Study of the equilibrium saddle point equation}

Here we carry out a detailed analysis of Eq. (\ref{saddle-eq}). The
equilibrium solution $z_s$ can be readily found in some simple limits. For $%
\mu \rightarrow \infty $ one can neglect in Eq. (\ref{saddle-eq}) the term $%
\frac 1{z+2\beta \mu }$ so to recover the case of the spherical SK model 
\cite{KTJ} with solution $z_s=1+\beta ^2$ for $T>1$. In the limit $%
T\rightarrow T_c\left( \mu \right) ^{+}$ we obtain at the leading order $z_s-%
\frac 2{T_c\left( \mu \right) }\simeq \frac 1{T_c\left( \mu \right) ^3}%
\left( \frac T{T_c\left( \mu \right) }-1\right) ^2$ with $\mu \neq -1$.
Finally, a zero temperature expansion gives $z_s\simeq -2\beta \mu +1+\frac{%
-\mu -\sqrt{\mu ^2-1}}\beta $ with $\mu \leq -1$.

Now we study Eq. (\ref{saddle-eq}) for $z$ varying in the complex plane,
since this is required during the discussion of the dynamics. Eq. (\ref
{saddle-eq}) is thus equivalent to: 
\begin{equation}
\left\{ 
\begin{array}{l}
\left( z-2\beta \right) \left( z+2\beta \right) =f\left( z\right) ^2\text{%
,\quad }f\left( z\right) =z-2\beta ^2\left( 1-\frac 1{z+2\beta \mu }\right) 
\\ 
\arg \left( z-2\beta \right) +\arg \left( z+2\beta \right) =2\arg f\left(
z\right) 
\end{array}
\right.   \label{saddle-eq b}
\end{equation}
with $\arg z\epsilon \left[ 0,2\pi \right[ $. The first of Eqs. (\ref
{saddle-eq b}) gives rise to an equation of third degree: 
\begin{equation}
C\left( z\right) =z^3-A_2z+A_1z-A_0=\left( z-a_1\right) \left( z-a_2\right)
\left( z-a_3\right) =0  \label{cubic}
\end{equation}
where 
\begin{equation}
\begin{array}{l}
A_2=a_1+a_2+a_3=2+\beta ^2-4\beta \mu  \\ 
A_1=a_1a_2+a_1a_3+a_2a_3=4\left( \beta \mu \right) ^2-6\beta \mu -4\beta
^3\mu +2\beta ^2 \\ 
A_0=a_1a_2a_3=\beta ^2\left( 1-2\beta \mu \right) ^2+4\beta ^2\mu ^2>0
\end{array}
\end{equation}
Furthermore one can find the following identity: 
\begin{equation}
\left( a_1-2\beta \right) \left( a_2-2\beta \right) \left( a_3-2\beta
\right) =\left[ 2\beta \left( \beta +\beta \mu \right) -2\left( \beta +\beta
\mu \right) -\beta \right] ^2\geq 0  \label{rel.roots}
\end{equation}
The previous relations can be used to get informations about the $a_k$
corresponding to a given choice of $T$ and $\mu $. For example, from the
zero temperature expansion of $z_s$ valid for $\mu \leq -1$, we obtain those
of $a_2,a_3$: $a_2\simeq -2\beta \mu +1+\frac{-\mu +\sqrt{\mu ^2-1}}\beta $, 
$a_3\simeq \beta ^2+\frac{2\mu }\beta $. However, in order to carry out a complete analysis of the possible $a_{k}$, we
proceed solving (\ref{cubic}) numerically. Let us now give the results of
our  numerical calculations. The plane $\mu -T$ can be divided in various
regions, as shown in Fig. 8. We have:

\begin{itemize}
\item  $a_1,a_2,a_3$ real with $a_1,a_2,a_3>2\beta $ in the regions $A_i$ $%
\left( i=1,2,3,4\right) $

\item  $a_1$ real, $a_2,a_3$ complex with $a_1>2\beta $ in the regions $B_i$ 
$\left( i=1,2\right) $

\item  $a_1,a_2,a_3$ real with $a_1>2\beta $ and $a_2,a_3<-2\beta $ in the
regions $C_i$ $\left( i=1,2,3,4\right) $
\end{itemize}

Furthermore, in the point $Q=\left( -\sqrt{2}/2,\sqrt{2}/4\right) $ the
three roots coincide: $a_1=a_2=a_3=6$. Along the lines $l_i$ $\left(
i=1,2,3,5,6\right) $ one has $a_2=a_3$ while for $l_4$ and in the zero
temperature limit with $\mu \leq -1$: $a_1=a_2$. In order to be solution of
Eq. (\ref{saddle-eq}) one $a_k$ has to satisfy also the second of Eqs. (\ref
{saddle-eq b}). Here is the list of such solutions in the various regions of
the phase diagram: 
\begin{equation}
\begin{tabular}{|l|l|}
\hline
{\bf Regions} & {\bf Solutions of Eq. (\ref{saddle-eq})} \\ \hline
$C_3$ & $z_s>2\beta ,-2\beta \mu \qquad \quad a_2,a_3<-2\beta ,-2\beta \mu $
\\ \hline
$l_2$ & $z_s>2\beta ,-2\beta \mu \quad \qquad a_2=a_3<-2\beta ,-2\beta \mu $
\\ \hline
$A_1$ & $z_s>2\beta ,-2\beta \mu \quad \qquad 2\beta <a_2<-2\beta \mu $ \\ 
\hline
$C_1$ & $z_s>2\beta ,-2\beta \mu \quad \qquad a_2<-2\beta ,-2\beta \mu $ \\ 
\hline
$A_2,A_3,B_1,l_1,l_5\qquad $ & $z_s>2\beta ,-2\beta \mu $ \\ \hline
$C_4$ & $a_2,a_3<-2\beta ,-2\beta \mu $ \\ \hline
$l_3$ & $a_2=a_3<-2\beta ,-2\beta \mu $ \\ \hline
$C_2$ & $a_2<-2\beta ,-2\beta \mu $ \\ \hline
$A_4,B_2,l_4,l_6$ & $none$ \\ \hline
\end{tabular}
\label{table}
\end{equation}
\begin{figure}[h]
\begin{center}
\epsfxsize=9cm
\epsffile{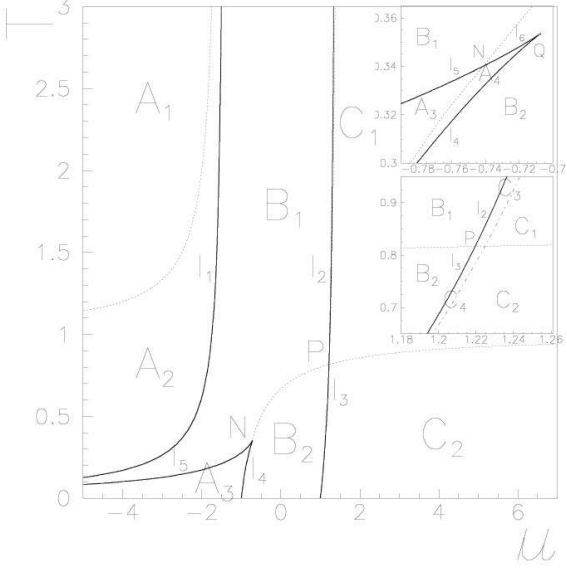}
\caption{ The various regions of the plane $\mu -T$ with respect to the roots 
$a_k$\ $\left( k=1,2,3\right) $. The characteristics of the roots in the
various regions are described in the text. The solid lines ($l_i$\ $%
i=1,...,6 $) are defind by the vanishing of the discriminant of (\ref{cubic}%
), namely $\frac 14\left( -\frac 2{27}A_2+\frac 13A_1A_2-A_0\right) ^2+\frac %
1{27}\left( A_1-\frac 13A_2^2\right) ^3=0$. The dotted lines are given by
Eq. (\ref{crit.line}) for $\mu <-1.5$ and $\mu \geq -1$. Finally, the
dash-dotted line is given by $T=\frac{1-\mu }{\mu -1.5}$ for $1\leq \mu <1.5$%
. The two \emph{insets} show an enlargement of the regions around the point $%
N\simeq \left( -0.74,0.34\right) $ and $P\simeq \left( 1.22,0.82\right) $.
Notice in particular the existence of the very narrow regions $A_4,C_3,C_4$.}
\end{center}
\end{figure}
\section{Computation of $\Gamma $ and of the density in the case of degenere
roots}

Here we discuss briefly the computation of the functions $\Gamma \left(
t\right) $ and $d\left( t\right) $ when the third degree polynomial $C\left(
z\right) $ has no distinct roots (lines $l_i$ in Fig. 8). Since the
equilibrium solution $z_s$ is never degenere, we do not find qualitative
differences with respect to the no degenere case; in particular the long time behaviours
of $\Gamma $ and $d$ are unchanged.

Firstly we consider the case of one doubly degenere root; let it be for example $a_2=a_3$ and thus $C\left( z\right) =\left(
z-a_1\right) \left( z-a_2\right) ^2$. Computing the inverse transform of (%
\ref{ga-lptr}), Eq. (\ref{ga ex1}) has to be replaced by: 
\begin{eqnarray}
\Gamma \left( t\right) &=&\phi _{11}^P\left( a_1\right) e^{\left( a_1/\beta
\right) t}-\beta \phi _{11}^Q\left( a_1\right) \int_0^te^{\left( a_1/\beta
\right) \left( t-u\right) }\frac{I_1\left( 2u\right) }udu  \nonumber \\
&&+\phi _{21}^P\left( a_2\right) \frac t\beta e^{\left( a_2/\beta \right)
t}-\beta \phi _{21}^Q\left( a_2\right) \int_0^te^{\left( a_2/\beta \right)
\left( t-u\right) }\frac{\left( t-u\right) }\beta \frac{I_1\left( 2u\right) }%
udu  \nonumber \\
&&+\phi _{22}^P\left( a_2\right) e^{\left( a_2/\beta \right) t}-\beta \phi
_{22}^Q\left( a_2\right) \int_0^te^{\left( a_2/\beta \right) \left(
t-u\right) }\frac{I_1\left( 2u\right) }udu
\end{eqnarray}
where $\phi _{11}^P\left( a_{1}\right) =\frac{P\left( a_1\right) }{C^{\prime
}\left( a_1\right) }$, $\phi _{11}^Q\left( a_{1}\right) =\frac{Q\left(
a_1\right) }{C^{\prime }\left( a_1\right) }$, $\phi _{2l}^P\left( a_{2}\right)
=\left. \frac{d^{l-1}}{dz^{l-1}}\frac{P\left( z\right) \left( z-a_2\right) ^2%
}{C\left( z\right) }\right| _{z=a_2}$ and $\phi _{2l}^Q\left( a_{2}\right)
=\left. \frac{d^{l-1}}{dz^{l-1}}\frac{Q\left( z\right) \left( z-a_2\right) ^2%
}{C\left( z\right) }\right| _{z=a_2}$ with $l=1,2\ $. Using the integral
representation of the modified Bessel function $I_1$ we find: 
\begin{eqnarray}
\Gamma \left( t\right) &=&e^{\left( a_1/\beta \right) t}S_1\left( a_1\right)
+e^{\left( a_2/\beta \right) t}\left[ \frac t\beta S_2\left( a_2\right)
+S_2^{\prime }\left( a_2\right) \right]  \nonumber \\
&&+e^{2t}\left[ \beta \phi _{11}^Q\left( a_1\right) J_{1,a_1/2\beta }\left(
2t\right) -\frac 12\phi _{21}^Q\left( a_2\right) J_{2,a_2/2\beta }\left(
2t\right) +\beta \phi _{22}^Q\left( a_2\right) J_{1,a_2/2\beta }\left( 2t\right)
\right]
\end{eqnarray}
with $S_1\left( a_1\right) $ given by (\ref{deS}); furthermore 
\begin{eqnarray}
S_2\left( a_2\right) &=&\phi _{21}^P\left( a_2\right) -\phi _{21}^Q\left(
a_2\right) \left( \frac{a_2-\sqrt{a_2^2-4\beta ^2}}2\right)  \nonumber \\
&=&\left\{ 
\begin{array}{ll}
\left[ \left( a_2+2\beta \mu \right) ^2-\beta ^2\left( a_2+2\beta \mu
-1\right) ^2\right] \frac 1{a_2-a_1}\quad & \text{if }a_2\text{ satisfies
Eq. (\ref{saddle-eq})} \\ 
\frac{C\left( a_2\right) }{a_2-a_1}=0 & \text{else}
\end{array}
\right.
\end{eqnarray}
and the integrals $J_{k,c}\left( t\right) $ for $k=1,2$ are defined by 
\begin{equation}
J_{k,c}\left( t\right) =\frac 1\pi \int_{-1}^1\frac{\sqrt{1-x^2}}{\left(
c-x\right) ^k}e^{\left( x-1\right) t}dx  \label{geik}
\end{equation}
Notice that the integral $J_{1,c}\left( t\right) $ coincides with that
defined by Eq. (\ref{I(t)}); furthermore $J_{2,c}\left( t\right) $ is
related to $J_{1,c}\left( t\right) $ by the relation $J_{2,c}\left( t\right)
=-\frac \partial {\partial c}J_{1,c}\left( t\right) .$ If $a_2$ does not
satisfy Eq. (\ref{saddle-eq}), the whole coefficient of $e^{\left( a_2/\beta
\right) t}$ vanishes since $C\left( a_2\right) =C^{\prime }\left( a_2\right)
=0$. The density is found to be: 
\begin{eqnarray}
d\left( t\right) &=&1-\frac T{2\Gamma \left( t\right) }\left\{ \frac{S\left(
a_1\right) }{a_1/2\beta +\mu }\left( e^{\left( a_1/\beta \right) t}-e^{-2\mu
t}\right) +\frac{S_2\left( a_2\right) }{a_2/2\beta +\mu }\left[ \frac t\beta
e^{\left( a_2/\beta \right) t}-\frac{e^{\left( a_2/\beta \right) t}-e^{-2\mu
t}}{2\beta \left( a_2/2\beta +\mu \right) }\right] \right.  \nonumber \\
&&+\frac{S_2^{\prime }\left( a_2\right) }{a_2/2\beta +\mu }\left( e^{\left(
a_2/\beta \right) t}-e^{-2\mu t}\right) +\frac{\phi _{11}^Q\left( a_1\right)
\beta }{a_1/2\beta +\mu }\left[ e^{2t}\left( J_{1,a_1/2\beta }\left(
2t\right) -J_{1,-\mu }\left( 2t\right) \right) -e^{-2\mu t}\left( \widetilde{%
J}_{1,a_1/2\beta }-\widetilde{J}_{1,-\mu }\right) \right]  \nonumber \\
&&-\frac{\phi _{21}^Q\left( a_2\right) }{2\left( a_2/2\beta +\mu
\right) }\left[ e^{2t}\left( J_{2,a_2/2\beta }\left( 2t\right) +\frac{%
J_{1,a_2/2\beta }\left( 2t\right) -J_{1,-\mu }\left( 2t\right) }{a_2/2\beta
+\mu }\right) +e^{-2\mu t}\left( \widetilde{J}_{2,a_2/2\beta }+\frac{%
\widetilde{J}_{1,a_2/2\beta }-\widetilde{J}_{1,-\mu }}{a_2/2\beta +\mu }%
\right) \right]  \nonumber \\
&&\left. +\frac{\phi _{22}^Q\left( a_2\right) \beta }{a_2/2\beta +\mu }%
\left[ e^{2t}\left( J_{1,a_2/2\beta }\left( 2t\right) -J_{1,-\mu }\left(
2t\right) \right) -e^{-2\mu t}\left( \widetilde{J}_{1,a_2/2\beta }-%
\widetilde{J}_{1,-\mu }\right) \right] \right\}
\end{eqnarray}
where $\widetilde{J}_{k,c}=J_{k,c}\left( 0\right) $; one has in particular $%
\widetilde{J}_{1,c}=c-\sqrt{c^2-1}$. We note that the long time behaviour of 
$J_{k,c}\left( t\right) $ is given by $J_{k,c}\left( t\right) \simeq \frac 1{%
\sqrt{2\pi t^3}\left( c-1\right) ^k}$; since it can be shown $\frac{\beta
^2\phi _{11}^Q\left( a_1\right) }{a_1-2\beta }-\frac{\beta ^2\phi
_{21}^Q\left( a_2\right) }{\left( a_2-2\beta \right) ^2}+\frac{\beta ^2\phi
_{22}^Q\left( a_2\right) }{a_2-2\beta }=\frac d{q_{EA}^2}$, one recovers the
final results of the Eqs. (\ref{ga-tltc},b), valid in the spin glass phase.

Finally, let us consider the case $a_1=a_2=a_3$ (point Q in Fig. 8): $%
C\left( z\right) =\left( z-a_1\right) ^3$. One has: 
\begin{equation}
\Gamma \left( t\right) =\sum_{l=1}^3\left[ \frac{\phi _{3l}^P\left(
a_1\right) }{\left( 3-l\right) !\left( l-1\right) !}\left( \frac t\beta
\right) ^{3-l}+\frac{\beta \phi _{3l}^Q\left( a_1\right) }{\left( 3-l\right)
!\left( l-1\right) !}\int_0^te^{\left( a_1/\beta \right) \left( t-u\right)
}\left( \frac{t-u}\beta \right) ^{3-l}\frac{I_1\left( 2u\right) }udu\right]
\end{equation}
where $\phi _{3l}^P\left( a_1\right) =\left. \frac{d^{l-1}}{dz^{l-1}}P\left(
z\right) \right| _{z=a_1}$, $\phi _{3l}^Q\left( a_1\right) =\left. \frac{%
d^{l-1}}{dz^{l-1}}Q\left( z\right) \right| _{z=a_1}\ l=1,2,3$. Then, using
the integral representation of $I_1$, one gets: 
\begin{eqnarray}
\Gamma \left( t\right) &=&e^{\left( a_1/\beta \right) t}\left[ \frac 12%
\left( \frac t\beta \right) ^2S\left( a_1\right) +\left( \frac t\beta
\right) S^{\prime }\left( a_1\right) +\frac 12S^{\prime \prime }\left(
a_1\right) \right]  \nonumber \\
&&+e^{2t}\left[ \frac{\phi _{31}^Q\left( a_1\right) }{4\beta }%
J_{3,a_1/2\beta }\left( 2t\right) -\frac 12\phi _{32}^Q\left( a_1\right)
J_{2,a_1/2\beta }\left( 2t\right) +\frac 12\beta \phi _{33}^Q\left(
a_2\right) J_{1,a_1/2\beta }\left( 2t\right) \right]
\end{eqnarray}
where 
\begin{eqnarray}
S\left( a_1\right) &=&\phi _{31}^P\left( a_1\right) -\phi _{31}^Q\left(
a_1\right) \left( \frac{a_1-\sqrt{a_1^2-4\beta ^2}}2\right)  \nonumber \\
&=&\left\{ 
\begin{array}{ll}
\left( a_1+2\beta \mu \right) ^2-\beta ^2\left( a_1+2\beta \mu -1\right)
^2\quad & \text{if }a_1\text{ satisfies Eq. (\ref{saddle-eq})} \\ 
C\left( a_1\right) =0 & \text{else}
\end{array}
\right.
\end{eqnarray}
and the integrals $J_{k,c}\left( t\right) $ are defined by (\ref{geik}) for $%
k=1,2,3$. In particular one has $J_{3,c}\left( t\right) =\frac 12\frac{%
\partial ^2}{\partial c^2}J_{1,c}\left( t\right) $. The coefficient of $%
e^{\left( a_1/\beta \right) t}$ vanishes since $C\left( a_1\right)
=C^{\prime }\left( a_1\right) =C^{\prime \prime }\left( a_1\right) =0$.\-\-
The density is 
\begin{eqnarray}
d\left( t\right) &=&1-\frac T{2\Gamma \left( t\right) }\left\{ \frac{\phi
_{31}^Q\left( a_1\right) }{4\beta \left( a_1/2\beta +\mu \right) }\left[
e^{2t}\left( J_{3,a_1/2\beta }\left( 2t\right) +\frac{J_{2,a_1/2\beta
}\left( 2t\right) }{a_1/2\beta +\mu }+\frac{J_{1,a_1/2\beta }\left(
2t\right) -J_{1,-\mu }\left( 2t\right) }{\left( a_1/2\beta +\mu \right) ^2}%
\right) \right. \right.  \nonumber \\
&&\left. +e^{-2\mu t}\left( \widetilde{J}_{3,a_1/2\beta }+\frac{\widetilde{J}%
_{2,a_1/2\beta }}{a_1/2\beta +\mu }+\frac{\widetilde{J}_{1,a_1/2\beta }-%
\widetilde{J}_{1,-\mu }}{\left( a_1/2\beta +\mu \right) ^2}\right) \right] 
\nonumber \\
&&-\frac{\phi _{32}^Q\left( a_1\right) }{2\left( a_1/2\beta +\mu \right) }%
\left[ e^{2t}\left( J_{2,a_1/2\beta }\left( 2t\right) +\frac{J_{1,a_1/2\beta
}\left( 2t\right) -J_{1,-\mu }\left( 2t\right) }{a_1/2\beta +\mu }\right)
+e^{-2\mu t}\left( \widetilde{J}_{2,a_1/2\beta }+\frac{\widetilde{J}%
_{1,a_1/2\beta }-\widetilde{J}_{1,-\mu }}{a_1/2\beta +\mu }\right) \right] 
\nonumber \\
&&\left. +\frac{\phi _{33}^Q\left( a_1\right) \beta }{2\left( a_1/2\beta +\mu \right)}%
\left[ e^{2t}\left( J_{1,a_1/2\beta }\left( 2t\right) -J_{1,-\mu }\left(
2t\right) \right) -e^{-2\mu t}\left( \widetilde{J}_{1,a_1/2\beta }-%
\widetilde{J}_{1,-\mu }\right) \right] \right\}
\end{eqnarray}
where $\widetilde{J}_{k,c}=J_{k,c}\left( 0\right) $. From $J_{k,c}\left(
t\right) \simeq \frac 1{\sqrt{2\pi t^3}\left( c-1\right) ^k}$ and $\frac 12%
\frac{\beta ^2\phi _{33}^Q\left( a_1\right) }{a_1-2\beta }-\frac{\beta ^2\phi
_{32}^Q\left( a_1\right) }{\left( a_1-2\beta \right) ^2}+\frac{\beta ^2\phi
_{31}^Q\left( a_1\right) }{\left( a_1-2\beta \right) ^3}=\frac d{q_{EA}^2}$
one gets again the final results of Eqs. (\ref{ga-tltc},b), valid in the
spin glass phase.


\begin{references}
\bibitem{NO}  As static properties are concerned, the two Ising
lattice fields $s_i,n_i$ can be replaced by the single spin-1 field $\tau
_i=s_in_i=0,\pm 1$ apart from a simple rescaling of $\mu $. The change of variables in Eq. (\ref{cv}) is rewritten in terms of $\tau_i$ as $\tau_i=\left( s_{1i}+s_{2i}\right) /2$.
\bibitem{MCT}  Traditionally, schematic MCT deals with a one time correlator
which, in our framework, is to be identified by $C_{nn}$ at equilibrium,
normalized by its equal time value: $\phi _{MCT}\left( t-t^{\prime }\right) =%
\frac{C_{nn}\left( t-t^{\prime }\right) }{C_{nn}\left( 0\right) }$.
\end{references}
\end{document}